\documentclass[fleqn,twoside]{article}
\usepackage{espcrc2}

\usepackage{graphicx}
\usepackage[figuresright]{rotating}

\newcommand{\AmS}{{\protect\the\textfont2
  A\kern-.1667em\lower.5ex\hbox{M}\kern-.125emS}}

\def\etal{ {\em et al.}}

\def\be {\begin{equation}}
\def\ee {\end{equation}}
\def\barr{\begin{array}}
\def\earr{\end{array}}
\def\dis{\displaystyle}
\def\ra{\rightarrow}

\def\bra {\langle}
\def\ket{\rangle}

\def\l {\lambda}

\def\rpv {R_p\!\!\!\!\!\!/~~}

\def\bpetapk {B^{\pm}\rightarrow \eta' K^{\pm}}
\def\betak {B\rightarrow \eta K}
\def\betak0 {B^{0}\rightarrow \eta' K^{0}}

\def\betapkstr0 {B^{0}\rightarrow \eta' K^{*0}}
\def\bpetak0 {B^{0}\rightarrow \eta' K^{0}}
\def\bpphik {B^{\pm}\rightarrow \phi K^{\pm}}

\def\msnus {m_{\tilde\nu_{iL}}^2}
\def\msells {m_{\tilde e_{iL}}^2}

\def\lappeq{\mathrel{\rlap{\raise.5ex\hbox{$<$}}
                    {\lower.5ex\hbox{$\sim$}}}}

\title{Decays $B \to \eta^{\prime} K$ in $R$-parity violating supersymmetry}

\author{B. Dutta\address[MCSD]{Center For Theoretical Physics, Department of
                Physics, Texas A$\&$M University,\\ College Station,
                TX 77843-4242, USA}%
                \thanks{The work of B.D. was supported in part by National
                Science Foundation grant No. PHY-0070964.},
        C. S. Kim\address[MCSD]{Department of Physics and IPAP, Yonsei University,
                Seoul 120-479, Korea}\thanks{The work of C.S.K. was supported
                in part by  CHEP-SRC Program, Grant No. 20015-111-02-2
                and Grant No. R03-2001-00010 of the KOSEF,
                in part by BK21 Program and Grant No. 2001-042-D00022 of the KRF,
                and in part by Yonsei Research Fund, Project No. 2001-1-0057.}
        and
        Sechul Oh\addressmark
                \thanks{The work of S.O. was supported by BK21 Program and Grant
                No. 2001-042-D00022 of the KRF.}}

\begin{document}

\begin{abstract}
In light of the recent experimental data from $B$ factories,
We try to explain the large branching ratio (compared to the Standard Model
prediction) of the decay $B^{\pm}\to \eta' K^{\pm}$ in the context
of $R$-parity violating ($\rpv$) supersymmetry.
We investigate other observed $\eta^{(\prime)}$ modes and find that only two pairs
of $\rpv$ coupling can satisfy the requirements without affecting the other
 $B \to PP$ and $B \to VP$ decay modes except the mode $B \to \phi K$.
We also calculate the  CP asymmetry for the observed decay modes
affected by the new couplings.
\vspace{1pc}
\end{abstract}

\maketitle

\section{Introduction}

Among the $B\ra PP$ ($P$ denotes a pseudoscalar meson) decay modes, the
branching ratio for the decay $\bpetapk$ is observed to be still larger than that
expected  within the Standard Model (SM). The SM contribution is about $3\sigma$
smaller than the experimental world average (Fig.1).  Among the $B\ra VP$ ($V$
denotes a vector meson) decay modes, the decay $\bpphik$ has  been observed
recently and the experimentally observed BR for the decay
$B^{0}\rightarrow \eta K^{*0}$ has been  found to be $2\sigma$ larger than the
SM. The BR for the newly observed decay
$B^{\pm}\rightarrow \eta K^{*\pm}$ is also available.

In this work \cite{dko}, we address these large BR problems of
$B^{\pm (0)}\rightarrow \eta^{(\prime)} K^{\pm (*0)}$ systems using
$R$-parity violating ($\rpv$) supersymmetric (SUSY) theories.  The effects of
$\rpv$ couplings on $B$ decays have been investigated previously in the
literature\cite{dkc}.  Attempts were made to fit just the large BR for
$\bpetapk$\cite{dkc}. At present, we  have more results. Some of these results
are concerned with decay
 modes involving
$\eta^{(')}$
 and these modes are influenced by the same $\rpv$ coupling that affects
$\bpetapk$. For example, the decay modes
$B^{\pm}\rightarrow \eta K^{*\pm}$,  $B^{0}\rightarrow \eta K^{*0}$,
$B^{0}\rightarrow \eta' K^{0}$  are affected by the new couplings which cure the
large BR problem of
$\bpetapk$. Hence,  it is natural to investigate  these newly observed  decay
modes and try to see whether all the available data can  be explained. We also
need to be concerned about the other observed (not involving
$\eta^{(\prime)}$) $B\ra PP$ and $B\ra VP$ decay modes, which could be
influenced by these new couplings. Our effort is not to affect the other modes
as much as possible, since except for  $B\ra \eta^{(\prime)}K^{(*)}$ decay
modes, the other observed modes fit the available data well\cite{fit,do}.
Further, using the preferred values of different parameters (e.g., new couplings
etc.), we also make predictions for CP asymmetrey for these observed modes which
will be verified in the near future.

\section{$\rpv$ SUSY effects to decay amplitudes}

The $\rpv$ part of the amplitude of  $B^{\pm}\ra\eta'K^{\pm}$ decay is

\begin{eqnarray}
{\cal M}_{\eta' K}^{\l'} &=& \dis
       \left( d^R_{121} - d^L_{112} \right) \xi A_{\eta'}^u  \\
       &+&
          \left( d^L_{222} - d^R_{222} \right) \:
                 \left[  \frac{\bar m}{m_s}
                   \left(A_{\eta'}^s -A_{\eta'}^u\right)- \xi  A_{\eta'}^s\right]
        \nonumber \\
       &+&   \dis
        \left(d^L_{121} - d^R_{112} \right) \frac{\bar m}{m_d} A_{\eta'}^u
        \nonumber \\
       &+&
       u^R_{112} \left[\xi A_{\eta'}^u - {2 m_K^2 A_K \over (m_s+m_u) (m_b-m_u)}
                 \right],   \nonumber
\label{eta1k}
\end{eqnarray}
where
\begin{eqnarray}
 d^R_{jkn} &=& \dis
      \sum_{i=1}^3 {\l'_{ijk}\l'^{\ast}_{in3}\over 8\msnus},  \nonumber \\
 d^L_{jkn} &=& \dis \sum_{i=1}^3 {\l'_{i3k}\l'^{\ast}_{inj}\over
8\msnus}, ~~(j,k,n=1,2)  \nonumber \\
 u^R_{jkn} &=& \dis \sum_{i=1}^3
{\l'_{ijn}\l'^{\ast}_{ik3}\over 8\msells}, ~~(j,k=1, \ n=2).
\end{eqnarray}
Here $\xi\equiv 1/N_c$ ($N_c$ denotes the effective number of color),
$\bar m \equiv m^2_{\eta'} / (m_b - m_s)$ and
\[
\barr{rcl} A_{M_1} & = & \bra M_2|J_b^\mu|B\ket \; \bra M_1|J_{l\mu}|0\ket.
\earr
\]
 $J$ stands for quark currents and the subscripts $b$ and
$l$ indicate whether the current involves a $b$ quark or only the light quarks.
Analogous expressions hold for $B^{\pm}\ra\eta K^{\pm}$ where  we have to replace
$A^u_{\eta^{\prime}}$ by $A^u_{\eta}$,
 $A^s_{\eta^{\prime}}$ by $A^s_{\eta}$ and $m_{\eta^{\prime}}$ by $m_{\eta}$.
Replacing a pseudoscalar meson by a vector meson, we
 also get similar expressions for the amplitudes of $B^{\pm(0)}\ra\eta'
K^{*\pm(0)}$ modes. The $\rpv$ part of the amplitude of
$B\ra\phi K$ decay mode involves   only $d^L_{222}$ and  $d^R_{222}$.

\be
\barr{rcl}  {\cal M^{\l'}_{\phi K}} &=& \dis

          \left( d^L_{222} + d^R_{222} \right) \:
                 \left[  \xi A_{\phi}\right],

\earr
\label{phik}
\ee
where $A_{\phi}=\bra K|J_b^\mu|B\ket \;\bra \phi|J_{l\mu}|0\ket$.

\section{Results}

We use the effective Hamiltonian and the effective Wilson coefficients
given in Ref. \cite{desh}.

\begin{table*}[htb]
\caption{{\bf Case 1:} for $|\lambda'| =0.06$ and $\xi=0.25$~.}
\label{table:1}
\newcommand{\m}{\hphantom{$-$}}
\newcommand{\cc}[1]{\multicolumn{1}{c}{#1}}
\renewcommand{\tabcolsep}{0.5pc} 
\renewcommand{\arraystretch}{0.6} 
\begin{tabular}{@{}lllllllll}
\hline
 & $\delta =0$, &
 & $\delta = 15^0$&  &$\delta = 0$, &
 & $\delta = 55^0$&\\
 & $\gamma =110^0$, &
 & $\gamma =110^0$&& $\gamma =80^0$, &
 & $\gamma =80^0$&\\ mode & ${\cal B} \times 10^{6}$ & ${\cal A}_{CP}$

& ${\cal B} \times 10^{6}$ & ${\cal A}_{CP}$& ${\cal B} \times 10^{6}$ & ${\cal
A}_{CP}$

& ${\cal B} \times 10^{6}$ & ${\cal A}_{CP}$
\\ \hline

$B^+ \to \eta^{\prime} K^+$ & 68.9 & 0.01 & 68.3 & 0.04 & 82.1 & 0.01 & 68.3 &
0.11
\\ $B^+  \to \eta K^{*+}$ &  36.4 & 0.03 & 36.4 & 0.04 & 36.5 & 0.03 & 32.7 &
0.09
\\ $B^0  \to \eta' K^0$ & 88.3 & 0.00 & 86.8 & 0.03 & 110.2 & 0.00 & 87.1 & 0.12
\\ $B^0  \to \eta K^{*0}$ & 14.0 & $-0.39$ & 14.6 & $-0.42$& 14.8 & $-0.28$ &
20.4 & $-0.56$
\\ $B^+  \to \phi  K^+$ & 7.11 & 0.00 & 6.97 & 0.04& 7.10 & 0.00 & 5.76 & 0.14
\\ \hline
\end{tabular}
\end{table*}
\begin{table*}[htb]
\caption{{\bf Case 2:} for $|\lambda'| =0.052$ and $\xi=0.45$~.}
\label{table:2}
\newcommand{\m}{\hphantom{$-$}}
\newcommand{\cc}[1]{\multicolumn{1}{c}{#1}}
\renewcommand{\tabcolsep}{0.7pc} 
\renewcommand{\arraystretch}{0.6} 
\begin{tabular}{@{}lllllllll}
\hline
 & $\delta =0$ & & $\delta =20^0$ &
\\ mode & ${\cal B} \times 10^{6}$ & ${\cal A}_{CP}$ & ${\cal B} \times 10^{6}$
& ${\cal A}_{CP}$
\\ \hline
$B^+ \to \eta^{\prime} K^+$ & 69.3 & 0.01 & 68.0 & 0.05
\\ $B^+  \to \eta K^{*+}$ &  27.9 & 0.04 & 27.8 & 0.05
\\ $B^0  \to \eta' K^0$ & 107.4 & 0.00 & 104.5 & 0.05
\\ $B^0  \to \eta K^{*0}$ & 20.5 & $-0.71$ & 21.1 & $-0.72$
\\ $B^+  \to \phi  K^+$ & 6.56 & 0.00 & 6.56 & 0.00
\\ \hline
\end{tabular}
\end{table*}

We first try to explain the large branching ratio of
$B^{\pm}\ra\eta'K^{\pm}$. The observed BR for this  mode in three different
experiments are\cite{cleo4,belle2,babar4}
\begin{eqnarray}
{\mathcal B}( B^{\pm}\ra\eta'K^{\pm}) \times 10^6 =80^{+10}_{-9}\pm 7 ~ ({\rm CLEO}),
\nonumber \\ ~~70\pm 8\pm 5 ~({\rm BaBar}),~~79^{+12}_{-11}\pm 9 ~({\rm Belle}).
\end{eqnarray}
The three results are close and we use the world average of them :
(in $10^{-6}$) $75 \pm 7$. The maximum BR in SM that we find is $42 \times 10^{-6}$
(Fig. 1). In the $\rpv$ SUSY framework, we find that the positive values of
$d^R_{222}$ and negative values of
$d^L_{222}$ can increase the BR keeping most of the other $B\ra PP$ and
$B\ra VP$ modes unaffected. The other $\rpv$ combinations are either  not enough
to increase in the BR or affect too many other modes.
(An important role is played by   the $\l'_{32i}$ -type couplings, the constraints
on which are relatively weak.)
We divide our results into
two cases,\\\noindent {\bf Case 1}: we use only
$d^{R}_{222}$ (positive values) and \\\noindent {\bf Case 2}: we  use a
combination of
$d^{R}_{222}$ (positive values) and
$d^{L}_{222}$ (negative values).

\vspace{0.5cm}
\includegraphics[width=15pc]{fig1.epsf} \\
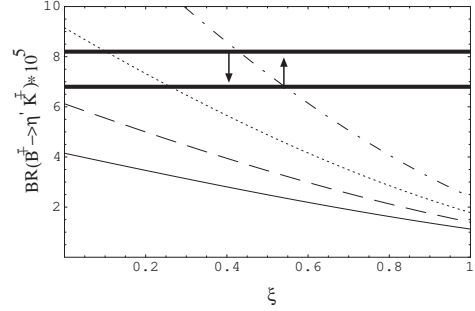
Figure 1. The BR for the decay  $B^{\pm}\rightarrow \eta'K^{\pm}$ vs
$\xi$. The solid line is for the SM. The dashed, dotted and dot-dashed  lines
correspond to $|\l'_{323}|=|\l'_{322}|=0.04$, 0.06, 0.08, respectively. The bold
solid lines indicate the experimental world average bound. \\ \\

Let us start with {\bf Case 1}.   We first discuss the case of $\gamma=110^0$.
In Fig. 1, we plot the BR for the decay
$B^{\pm}\ra\eta^{\prime} K^{\pm}$  as a function of
$\xi$. We have used  $|\l'_{323}|=|\l'_{322}|=0.04$, 0.06, 0.08 and $m_{\rm
susy}=200$ GeV. We take $d^R_{222}$ to be positive.
The large branching ratio can be explained for
$\l'\geq0.05$.

The ${\cal B}(B\rightarrow X_s \nu\nu)$ can put bound on
$\l'_{322}\l'^{\ast}_{323}$ in certain limits: $\l'\leq 0.07 $.

In another scenario we can use smaller value of $\gamma$, e.g., $\gamma=80^0$
to fit the $B\ra \eta^{(\prime)} K^{(*)}$ data.
In Table I, we present the BRs and the CP asymmetries for $B\ra
\eta^{(\prime)} K^{(*)}$ and $B \ra \phi K$ for different values of $\delta$
and $\gamma$.
Here the phase difference between $\lambda_{323}^{\prime}$
and $\lambda_{322}^{\prime}$, $\delta$, is defined by
\begin{equation}
\l'_{323} \l'^{\ast}_{322} =|\l'_{323}
\l'^{\ast}_{322}|e^{i\delta}.
\end{equation}
 The maximum values of $\delta$ allowed by the BR of
$B^{\pm} \ra\eta' K^{\pm}$ are $\delta=15^0$ for $\gamma=110^0$ and
$\delta=55^0$ for $\gamma=80^0$.   \\

\noindent{\bf Case 2}: We now use the combination of $d^R_{222}$ and $d^L_{222}$
with $\gamma=110^0$.  We assume $d^R_{222}=-d^L_{222}$.
 In this scenario, the $\rpv$ coupling part of the amplitude in $B\ra\phi K$
decay mode canceled exactly (Eq. \ref{phik}). (In fact, our solution still works when
the cancellation is incomplete by about 5$\%$.)
But we still have contributions to $B^{\pm}\ra\eta' K^{\pm}$ (Eq. \ref{eta1k})
and to increase the BR we choose
$d^R_{222}$ to be positive. There is no $\rpv$ contribution to the other $B\ra
PP$ and $B\ra VP$ modes in this case as well.

In Table II, we calculate the BRs and the CP asymmetries for $B\ra
\eta^{(\prime)} K^{(*)}$ and $B \ra \phi K$ for different values of $\delta$.
The maximum value of $\delta$ allowed by the
BR  for $B^{\pm} \ra\eta' K^{\pm}$ is $\delta=20^0$ for $\gamma=110^0$.

\section{Conclusion}

We have studied $B \ra\eta^{(\prime)} K^{(*)}$ modes in the
context of $\rpv$  SUSY theories.
We have found solutions for both large and small values of $\gamma=110^0,\,80^0$ and
two different values of $\xi\simeq 0.25,\,0.45$ for two different scenarios. For
our solutions, we need $|\l'| \sim 0.05-0.06$ for $m_{\rm susy}=200$ GeV.

\end{document}